\documentclass[fleqn,usenatbib]{mnras}

\usepackage{mathptmx}
\usepackage{ulem}

\usepackage{graphicx}	
\usepackage{amsmath}	
\usepackage{pdflscape}	

\usepackage[T1]{fontenc}

\DeclareRobustCommand{\VAN}[3]{#2}
\let\VANthebibliography\thebibliography
\def\thebibliography{\DeclareRobustCommand{\VAN}[3]{##3}\VANthebibliography}

\usepackage{amsmath}	
\usepackage{amssymb}	

\title{Microwave response to sunspot oscillations}

\author[Robert Sych et al.]{
Robert Sych,$^{1}$\thanks{E-mail: sych@iszf.irk.ru}
Alexander Altyntsev$^{1}$
\\
$^{1}$Institute of Solar-Terrestrial Physics SB RAS, Irkursk, Russia\\
}

\date{Received: September 2022}

\pubyear{2022}

\begin{document}
\label{firstpage}
\pagerange{\pageref{firstpage}--\pageref{lastpage}}
\maketitle

\begin{abstract}
We present the first observations of spatially resolved oscillation sources obtained with the Siberian Radioheliograph (SRH) at 3-6 GHz. We have found significant flux oscillations with periods of about 3, 5 and 13 minutes emitted from AR12833. The 3-minute periodicity dominates at higher frequencies. It was found that the apparent level of oscillations depends on the active region location on the disc, and scales down towards the limbs. The oscillations were studied in detail during one hour interval on 2021, June 19. We found that sources of 3-min oscillations were located above the umbra and their emission is extra-ordinary polarized.The 5 and 13-min periods were manifested in emission at lower frequencies, down to 2.8 GHz. Sources with 5-min periodicity were located near the umbra/penumbra boundary and in the pore region. Positions of sources with 13-min oscillations were different at 3.1 GHz and 4.7 GHz. We found consistency between spatial location of the oscillation sources in radio and UV in 171 \AA ~and 304 \AA. ~There is significant correlation of signals in two ranges. Time delays between microwave oscillations increase as the frequency decreases, which can be explained by upward propagation of periodic disturbances. The localization of oscillation sources is probably related to magnetic structures with different wave cutoff frequencies at different heights. The obtained results show that  SRH can provide the spatial resolved observation of the oscillations in the intensity and polarization channels in 3–6 GHz band.

\end{abstract}

\begin{keywords}
sunspot -- oscillations -- radio radiation -- magnetic fields -- UV radiation
\end{keywords}



\section{Introduction}

From observations in chromospheric lines, it is known that in emission from around the sunspots, one can see changes in the intensity and Doppler shift of velocities, which manifest as quasi-periodic oscillations (QPO) \citep{1992ASIC..375..261L}. These oscillations are directly related to the propagation of solar magnetohydrodynamic (MHD) waves and possibly play an important role in heating the corona and chromosphere \citep{1969SoPh....7..351B, 1972SoPh...27...71G}. 

In a wide range of heights, starting from the transition zone, observations in microwave emission are the important means for diagnostics of coronal structures above the sunspots. To detect components of harmonic emission, observations with high spatial resolution are required.
First observations with the small-base radio interferometer detected oscillations in the active region emission at 8.6 GHz with periods of $\sim$3, 5 and 7 min \citep{1984SoPh...93..301Z}. Further, the use of one-dimensional scans of the Sun obtained with the Siberian Solar Radio Telescope (SSRT, 5.7 GHz) \cite{1989IGAFS..87...113Z, 1992IGAFS..98...114Z} showed that oscillations are caused by changes in radio flux from local gyroresonance sources related to sunspots. A review of the properties of the gyroresonance mechanism in strong radio sources associated with sunspots has been provided in \cite{1997SoPh..174...31W}.

Important results on microwave QPOs were obtained using observations of the Nobeyama Radioheliograph at 17 GHz with up to 10\arcsec ~angular resolution. In \cite{1999SoPh..185..177G} it was found that nearly harmonic oscillations with periods of 120–220 s can be seen in polarized emission flux of sunspots. It was suggested that these oscillations can be explained by magnetohydrodynamic waves that change the size and temperature of the layer emitting due to the gyroresonance mechanism. The use of information on plasma density and temperature in the transition layer obtained from SOHO/SUMER EUV emission data, allowed \cite{2001ApJ...550.1113S} to confirm the assumption that microwave oscillations are caused by sound waves propagating upward. It was concluded that variations in radio brightness at 17 GHz within 3000K are caused by oscillations of density and temperature within the region of the third gyroresonance level with magnetic field of $\sim$2000 G. The 3-min QPOs was associated with the resonant excitation of the cutoff frequency in the region of temperature minimum. 

Investigation into spatial, temporal and phase features of sunspot oscillations at 17 GHz continued using the pixel wavelet filtration method (PWF analysis) in \cite{2008SoPh..248..395S}. The authors measured the dynamics and spatial distribution of power for different harmonics in the oscillation spectrum and found that the 3-min harmonic source is placed in the umbra, and the sources of 5-min component are mainly located on the umbra/penumbra boundary.

Development of extra-atmospheric observations of QPO sources in ultraviolet and extreme ultraviolet emission brought new opportunities to study their spatial structure. Observations with high spatial resolution show that the structures above sunspots are sharply inhomogeneous crosswise the magnetic field, at scales significantly smaller than the Nobeyama Radioheliograph spatial resolution. Thus, the apparent microwave emission sources result of convolution of the several compact sources with the beam of radioheligraph.  

Observations with the large radio interferometer VLA (spatial resolution of 5.7\arcsec x 5.3\arcsec ~at 5 GHz and 3.0\arcsec x 2.8\arcsec ~at 8.5 GHz) show that the observed sizes of 3-min QPO sources do not exceed few arcseconds, i.e. they are comparable to the diagram size \citep{2002A&A...386..658N}. Important result  is the conclusion on stability of positions, amplitude and phase of oscillation sources during many periods, which explains the possibility of observing them using instruments with relatively low spatial resolution. Modeling showed that the observed oscillations can be interpreted as the result of variations in position of the second/third level of gyrofrequency level  relative to the transition layer boundary. The observed levels of radio emission modulation did not exceed fractions of a percent and can be provided by variations in magnetic field with the amplitude of $\sim$ 40 G, or by 25 km changes in the height of the transition layer base. Another possible explanation might be variations in the direction of magnetic waveguides' vector within several degrees, along which disturbances propagate.

Consequently, observations of QPOs in the microwave range provide detailed information on sunspot oscillations in the region where surfaces of gyrofrequency levels intersect with the transition zone boundaries. In this regard, simultaneous microwave observations with novel radioheliographs (SRH, FASER, MUSER) at different frequencies, i.e. in different locations above the sunspot atmosphere, allows to determining the wave propagation velocity.

The purpose of this paper is to study the potential of observations of active region sunspot oscillations when testing the Siberian Radioheliograph (SRH) at several frequencies simultaneously in 3–6 GHz band. Data of SRH observations are compared with the images obtained with SDO/AIA spacecraft in UV 304 \AA ~and 171 \AA.

The paper is arranged as follows: in Section 1, we introduce the paper subject; in Section 2, we provide the observational data and processing methods; in Section 3, we describe the data analysis and obtained results; in Section 4, we discuss the processes of sunspot oscillations and structure of the active region; and in Section 5, we make conclusions concerning the obtained results.

\section{Observations and data processing}

We analyzed radio signal oscillations measured from the sunspot active region NOAA 12833 on June 14–24, 2021. At this time, a large symmetrical N-polarity sunspot sized $\sim$40\arcsec ~was passing across the solar disk. We used correlation curves and time cubes of solar images obtained during $\sim$10 hour daily observation with the SRH of at 3–6 GHz. T-shaped array comprised 107 equidistant antennas 3 m in diameter. Design and technical characteristics of SRH are described in \cite{2020STP.....6b..30A}.

To study oscillations in detail we have selected an interval on June 19, 2021 at 01:30–02:30 UT when the sunspot was near the solar center.
The sunspot-related radio source was centered at the initial time of 01:30 UT, its spatial motion being compensated due to the solar differential rotation. At this time, the sunspot was near the central meridian. SRH observations were  carried out at fixed frequencies of 2.8 GHz (10.7 cm), 3.1 GHz (9.7 cm), 3.4 GHz (8.8 cm), 3.9 GHz (7.7 cm), 4.7 GHz (6.4 cm), and 5.6 GHz (5.4 cm). The diagram size for these frequencies changed inversely to frequency, starting from 19.6” x 14.8” at 2.8 GHz with pixel size of 5\arcsec. Using visibility functions we synthesized the images and performed their brightness temperature calibration using the SRH remote access and data processing system, implemented under the leadership of Dr. Lesovoi S.V. Data were recorded in the intensity (I) and circular polarization (V) channels with the cadence of 1.32 s for correlation curves and 12 s for images. To compare the modulation levels in the active region and the quiet Sun region, we selected two 115” x 125” regions separated in latitude.

To calculate the change in 3-min oscillation power during the passage of NOAA 12833 active region across the solar disk on June 14–24, 2021, we used daily correlation curves at 02–04 UT. During this interval, the diagram effects on the correlation changes over time are minimal. To calculate the power of oscillations, we applied temporal wavelet analysis of variations in correlation coefficients and plotted the global wavelet spectrum in the period range of $\sim$ 3 min.

To compare the SRH data with those at other wavelengths, we obtained the UV intensity data cube from the Atmospheric Imaging Assembly (AIA) on board the Solar Dynamics Observatory (SDO) \citep{2012SoPh..275....3P} in the bandpass of 304 \AA ~(He II: chromosphere and transition region) and 171 \AA ~(Fe IX: quiet corona). The time cube of images was obtained using the SDO data preparation website~ \footnote{\url{http://jsoc.stanford.edu/ajax/lookdata.html}}. It allows obtaining images of the Sun for different wavelengths and in a given time interval. For data processing (centering and absolute calibration), we used the Solar Soft library. Also, we removed the differential rotation of the source. We used  300 images obtained with cadence of 12 s and spatial resolution of $\sim$ 1.2\arcsec.

For spectral analysis and detection of oscillation harmonics, we used wavelet analysis and constructed wavelet power and global spectra. To find the relationship between radio and UV data, we used correlation analysis, both in linear approximation and for different periods using wavelet cross-correlation. Combination of wavelet filtering and cross-correlation function was first proposed in \cite{1995CRASB.321..525N} to study solar activity. This method is purposed for simultaneous analysis of two signals in frequency and time domain. Its main advantage is that it allows studying how the parameters of two series (amplitude, intensity, correlation, phase and coherence) evolve over time.

\section{Results}

\begin{figure*}
\includegraphics[width=18 cm]{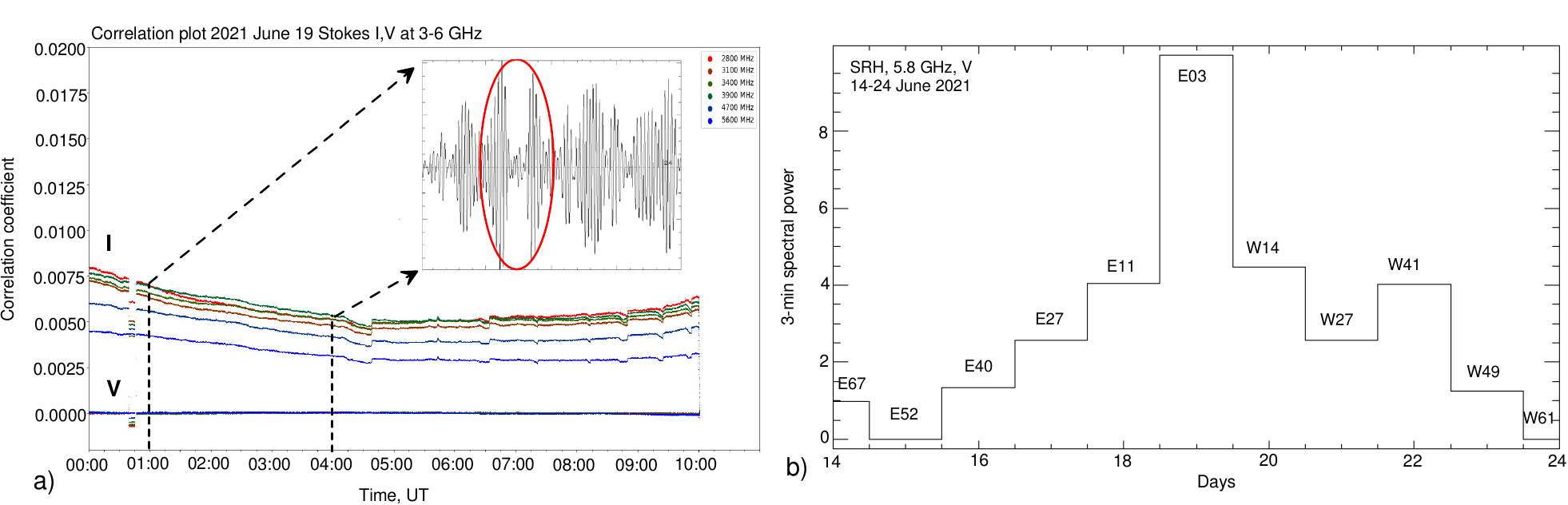}
\caption{(a) Microwave correlation curves obtained on June 19, 2021 at 3–6 GHz. Data in the intensity (I) and the circular polarization (V) channel are shown. Dashed vertical lines show the time interval with 3-min oscillations. Ellipse denotes the studied time interval at 01:30-02:30 UT. Different colors of curves correspond to different radio frequencies. (b) Spectral power of 3-min oscillations of correlation coefficients at 5.6 GHz during during the passage of NOAA 12833 sunspot active group passing across the solar disk.}
\label{1}
\end{figure*}

We used the SRH database of correlation curves and images for visual search of oscillations in microwave emission. Curves are plotted online at the SRH site \footnote{\url{https://badary.iszf.irk.ru/}}  and are available as FITS files. Figure~\ref{1}(a) shows an example of correlation curves for June 19, 2021. Different color profiles correspond to different frequencies of emission. Dashed vertical lines bound the interval of $\sim$3 hours when oscillations were observed.

\begin{figure}
\includegraphics[width=8.5 cm]{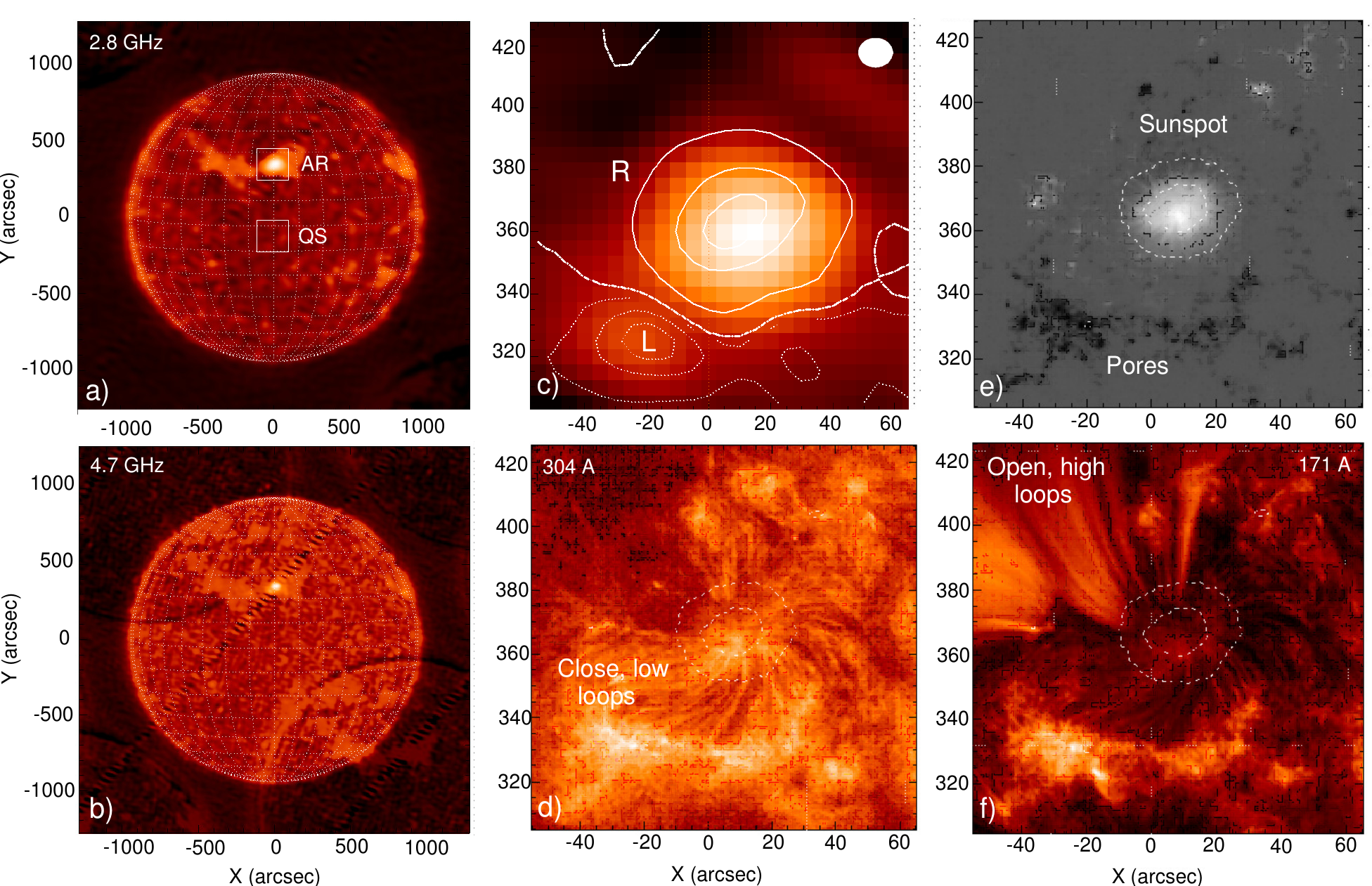}
\caption{Radio images of the Sun at 2.8 GHz (a) and 4.7 GHz (b) on June 19, 2021, 01:30 UT in the intensity channel. The studied NOAA 12833 active region (AR) and quiet Sun (QS) region are shown with white squares; (c) sunspot image in the intensity (background) and in the polarization (R and L as contours) channel at 4.7 GHz. Contour levels are 0.1, 0.5, and 0.9 of peak brightness. White ellipse shows the SRH diagram. Thick dashed line is the neutral line; (d) UV image of the active region in 304 \AA ~(SDO/AIA); (e) AR magnetogram (SDO/HMI). Contours show umbra and penumbra regions; (f) AR image in 171 \AA ~coronal line.}
\label{2}
\end{figure}

Figure~\ref{1}(b) shows the dynamics of power changes for 3-min oscillations during the passage of a unipolar N-polarity sunspot in NOAA 12833 active group across the solar disk on June 14–24, 2021. The sunspot longitude in heliographic coordinates is indicated for the day of measurement. As the sunspot was moving towards the center, the power of oscillations in the polarization channel increased by an order of magnitude and reached its maximum on June 19, 2021, near the central meridian. This increase is noted in the longitude range limited to      $\sim$40 degrees. There are no oscillations near the limbs or their amplitudes are low due to the projection effect. When the active region moved across the solar disk, one could notice the active region evolution as formation of a belt of small pores near the sunspot. These changes reach their maximum on June 19–20. Parallel to this, the sunspot area was growing.

To study the oscillations we selected one-hour interval at 01:30-02:30 UT on June 19, 2021 with intense oscillations, which is highlighted with red ellipse on filtered 3-min profile at 5.6 GHz in Fig.~\ref{1}(a). Emission oscillations have a low-frequency modulation in the form of wave trains. The duration and amplitude of train modulation vary over time, which points out that oscillations are non-stationary. It should be noted that this dynamics of sunspot oscillations was previously discussed in a number of works \citep{2006ApJ...647L..73H, 2007ApJ...655..624D, 2012A&A...539A..23S}.             

To analyze oscillations, sequences of the full solar disk images averaged over 10 s intervals were processed for each radio frequency. This made it possible to reduce high-frequency noise and increase the contrast of radio brightness in the images. Figures~\ref{2}(a,b) shows the example of the solar radio maps that were obtained on 01:30 UT at 2.8 and 4.7 GHz.

We can see that there is only one powerful radio source on the disc associated with the sunspot active region NOAA 12833. A white square with AR notation indicates its position. The quiet Sun (QS) region is also marked, which we used it to compare the oscillations between them.

Along with studies of the region in radio, we used time cubes of UV images obtained with the SDO/AIA space telescope at 304 \AA ~and 171 \AA ~at 01:30–02:30 UT. The magnetogram and optical image of the sunspot were obtained at the initial time point of 01:30 UT from SDO/HMI data.
 
The study area includes R (+) polarized sunspot and L (–) polarized pores (see Fig~\ref{2}c). Comparison with the magnetic field of the N-polarity sunspot indicates the extraordinary X-mode of emission. White ellipse depicts the SRH diagram. Solid contours indicate the sunspot region at 4.7 GHz. Its localization matches the N-polarity sunspot in magnetogram (see Fig.~\ref{2}e), where dashed line depicts the boundaries of umbra and penumbra. The maximum of the sunspot magnetic field is $\sim$3000 G. The source brightness temperature is typical for the corona making up $1.3x10^6$ K (2.8 GHz) and $1.9x10^6$ K (4.7 GHz). Polarization is higher at 4.7 GHz (12.1\%) compared to 5.2\% at 2.8 GHz.
 
South of the sunspot, there is a region with the opposite L-polarity, Fig.~\ref{2}(c), which coincides with dark pores of small angular sizes, whose negative field reaches $\sim$1000 G. Distinguished are two small polarized radio sources related to bright details in the form of a pore cluster. They are shown with thin dotted lines. Thick dashed line depicts the active region neutral line. Similar to the sunspot, emission in the pores has an extraordinary X-mode.

In the transition zone, ~above the solar surface, we can clearly see cold and dense formations shaped as fibers and loops. In Fig.~\ref{2}(d), we can see a series of low closed magnetic loops that are anchored in the sunspot center and come out with some twist, connecting to a bright region of the opposite polarity at the bottom. In 171 \AA ~coronal line (see Fig.~\ref{2}e), we also observe these low loops, but at these heights, open high loops become the brightest ones. As in the previous case, they are anchored in the umbra.

Spatial locations of magnetic loop structures show that outside the sunspot, magnetic field diverges asymmetrically. In the southern direction, one can observe low loop arcades closing on the nearest region of the opposite polarity. There exist open magnetic structures in the northern direction.
 
Variations in correlation curves are mainly associated with emission from compact regions of small angular size, where the signal correlation is maximal. In our case, we observed only one active region NOAA 12833 with a large symmetrical sunspot, without significant flare activity. It can be assumed that the oscillations seen during the observations are associated with 3-min oscillations in the sunspot that are well studied in different spectral ranges \citep{1986ApJ...301..992L, 2010ApJ...722..888B, 2006A&A...456..689T}. To verify this statement, we mapped the global signal variation at 01:30–01:40 UT for each pixel of the whole solar disk. We used time cube of synthesized images obtained at 4.7 GHz in the intensity channel, with 12 s cadence between the images. This allowed revealing the region with the highest level of variations.

\begin{figure}
\includegraphics[width=8.5 cm]{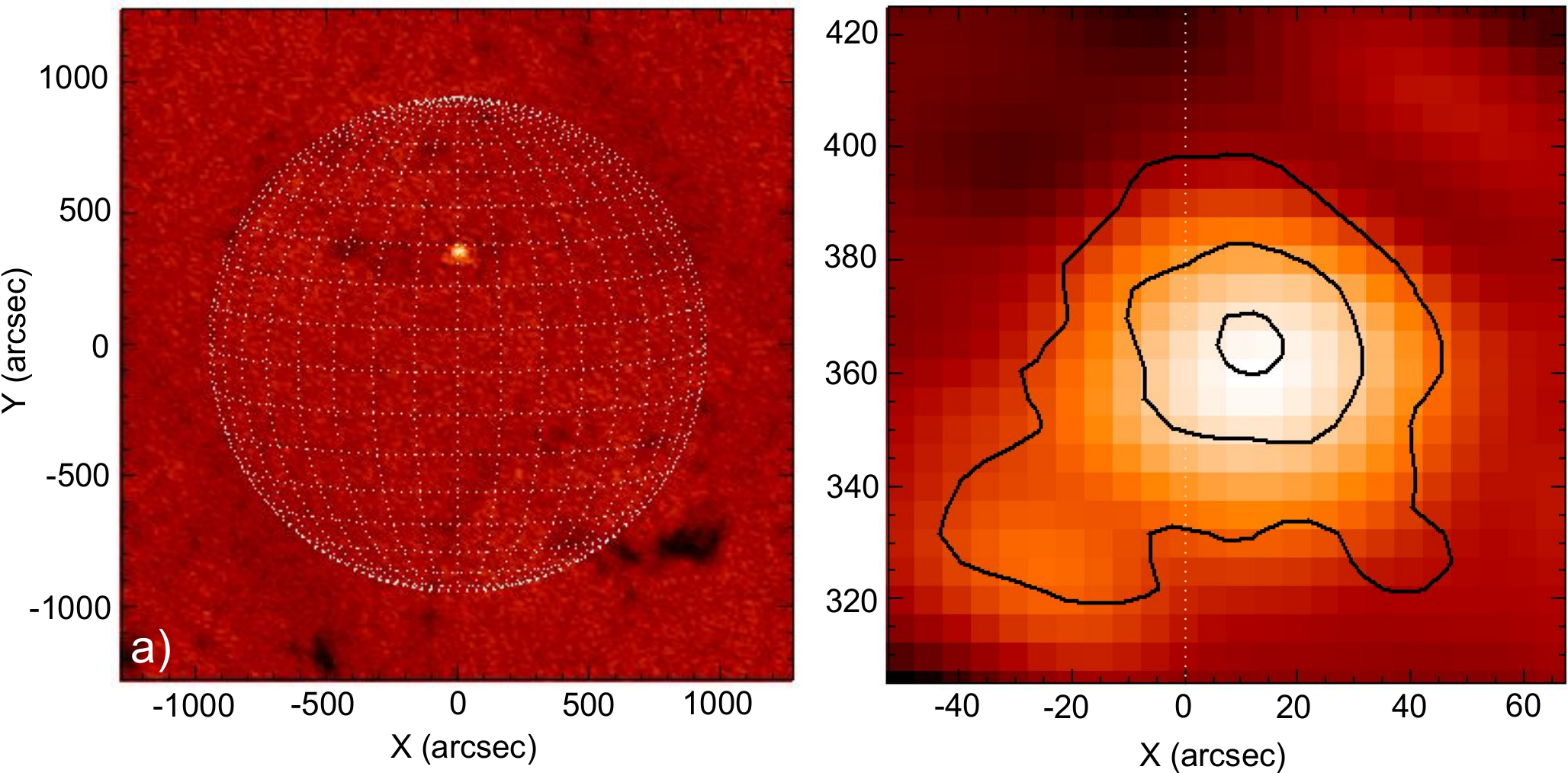}
\caption{Variation maps of the temporal cube of solar images at 01:30–01:40 UT in the intensity channel (a) and the studied active region NOAA 12833 (b, background) with superimposed contours of radio variation at 4.7 GHz. Contour levels are 0.1, 0.5, and 0.9 of the maximum variation.}
\label{3}
\end{figure}

In the variation map Fig.~\ref{3}(a) we can see only one compact source on the solar disk that stands out significantly in brightness compared to the quiet Sun. This source is connected with the NOAA 12833 sunspot active region. Superimposing the highest variation region in the form of contours on the radio source region as a background image showed their good spatial match (see Fig.~\ref{3}b). Variations involve the entire radio source, not only its central part.

\subsection{Temporal dynamics of oscillations}

\begin{figure}
\includegraphics[width=8.5 cm]{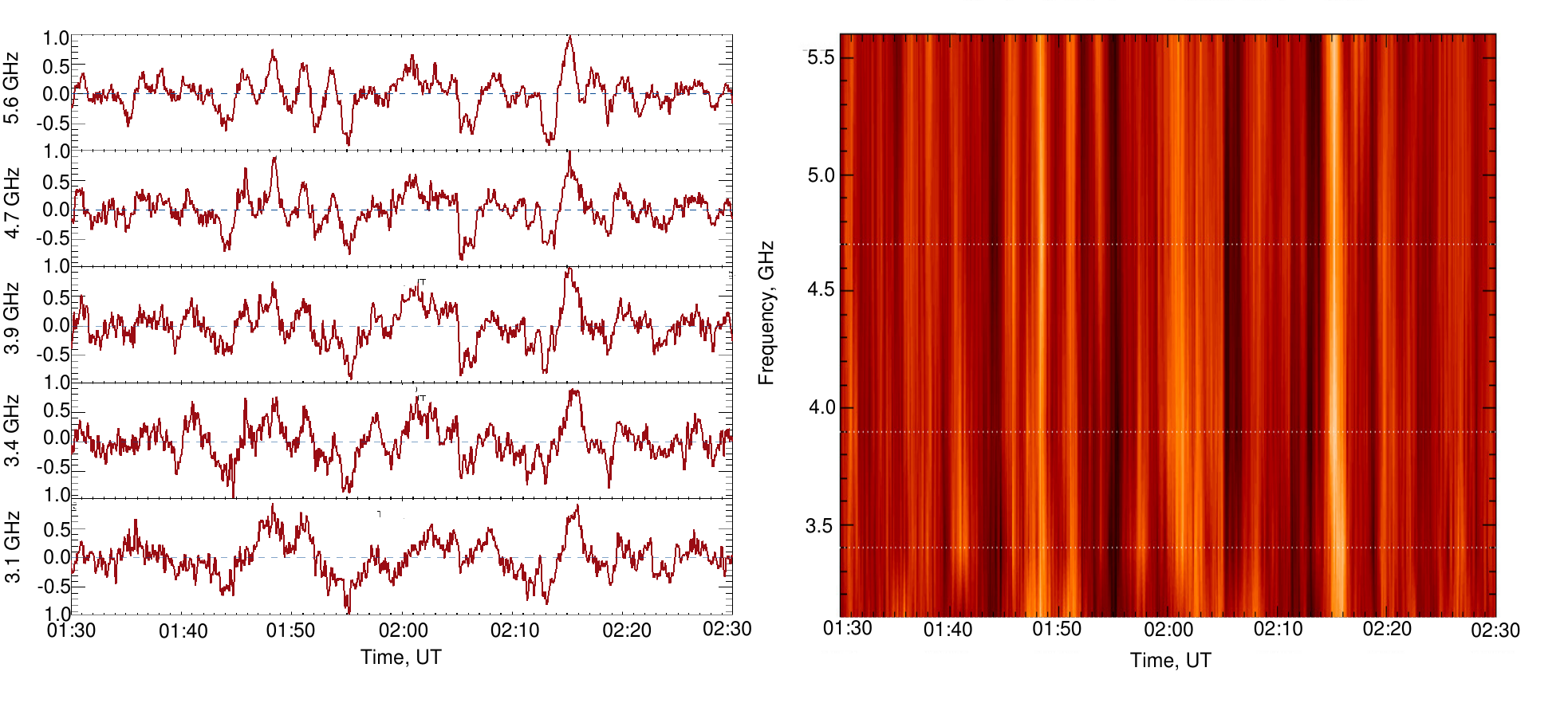}
\caption{Left panels: microwave correlation curves obtained with SRH on June 19, 2021 during one-hour observations at 01:30–02:30 UT. Curves for frequencies at 3.1, 3.4, 3.9, 4.7, and 5.6 GHz in the intensity channel are shown. Right panels: dynamic spectrum obtained from correlation curves for relevant channels. Horizontal dotted lines show their frequency location.}
\label{4}
\end{figure}

To investigate the revealed oscillations, we compared the correlation profiles for six frequencies  in 3-6 GHz frequency band and calculated the dynamic spectrum from them. The curves were obtained with the removed daily parabolic trend in the intensity channel (see Fig.~\ref{4}, left panels). Visually, significant variations are detected as the oscillation period increases from $\sim$3 to $\sim$13 min when the frequency of signal receiving gets lower. The signal-to-noise ratio grows with frequency.

Figure~\ref{4}, right panel demonstrates alternation of light and dark strips related to 3-min oscillations. Their intensity is modulated synchronously by oscillations with lower periods. Vertical strips are almost parallel to each other, i.e. the oscillations are phased at all frequencies shown as horizontal dotted lines. 3-min periodicity is most pronounced at 5.6 GHz. At low frequencies, its magnitude decreases.

To reveal phase difference of signals in a frequency band, we choose a train of 3-min oscillations at 01:40–02:00 UT when they have the highest power. We can see the increase in phase difference between 5.6 GHz reference frequency and lower frequencies. Thus, for pairs 5.6–4.7 GHz, 5.6–3.4 GHz and 5.6–2.8 GHz, the differences are 9.7, 33 and 65 degrees. This corresponds to time delays of 4, 14 and 28 s, taking the account of the average oscillation period of $\sim$2.6 min. It should be noted that during the development of the wave train for different frequency pairs, we could observe both constant phase difference and their changes over time. 

To compare the time parameters of SRH integral (correlation curves) and two-dimensional (image cube) data, we calculated modulation of radio flux from the sunspot and quiet regions in the intensity channel for 4.7 GHz, with the highest level of variation. Modulation was calculated in percent as follows: $I (t) - Aver/Aver$, where $I (t)$ represents oscillations of brightness temperature for SRH and $Aver$ is average of   $I(t)$. We co-aligned all the images to the initial time 01:30 UT and integrated brightness of the source inside half-maximum contour. For the quiet Sun, we integrated the emission from the entire selected frame. Then the obtained radio flux curves and correlations at the same frequency were compared to each other using the cross-correlation method.

\begin{figure}
\includegraphics[width=8.5 cm]{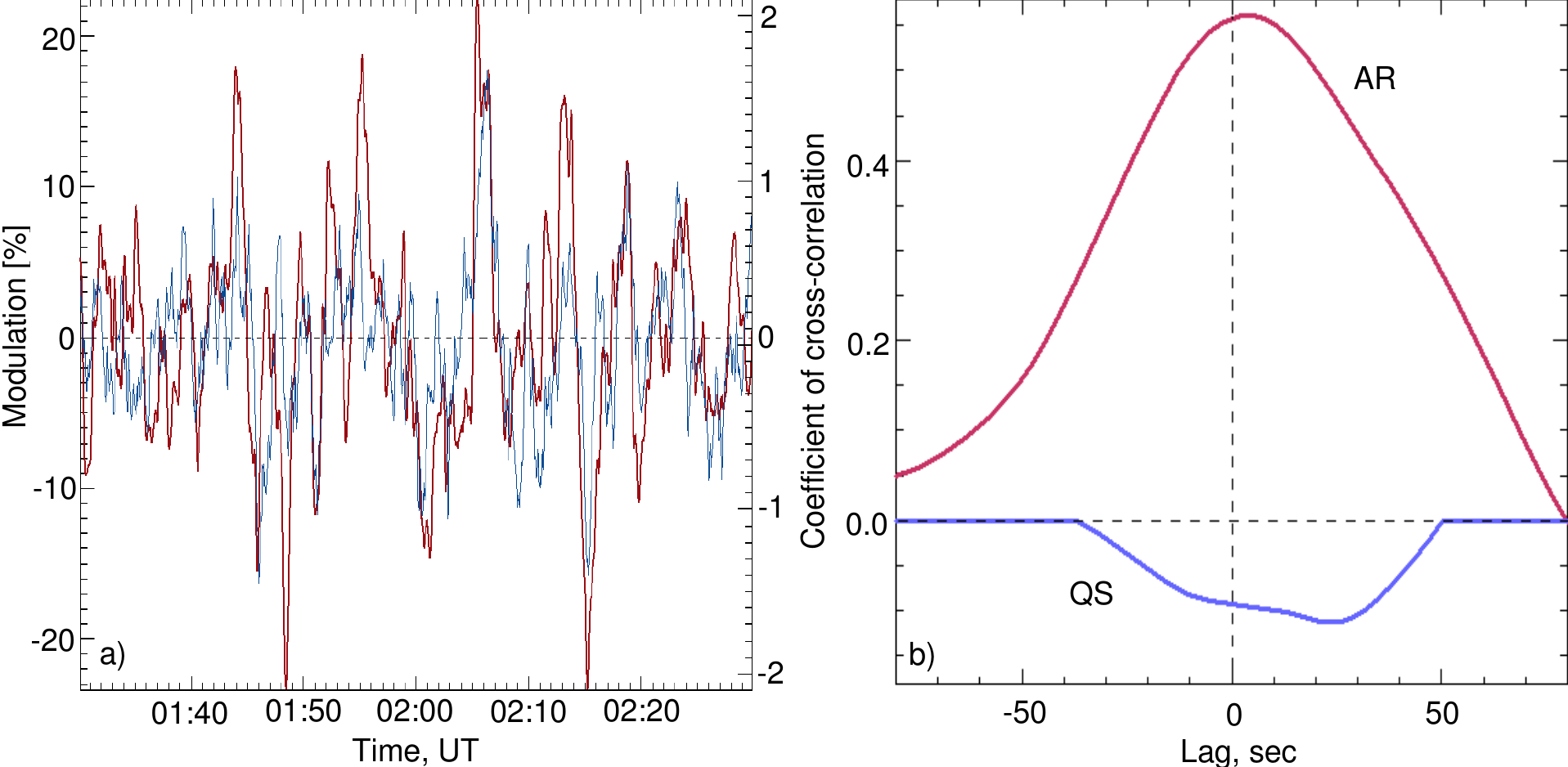}
\caption{(a) Modulation profiles of correlation curve (brown lines) and local source of radio flux (blue lines) in the intensity channel at 4.7 GHz (b). Cross-correlation spectra between correlation curve and radio fluxes of AR and QS.}
\label{5}
\end{figure}

\begin{figure*}
\includegraphics[width=18 cm]{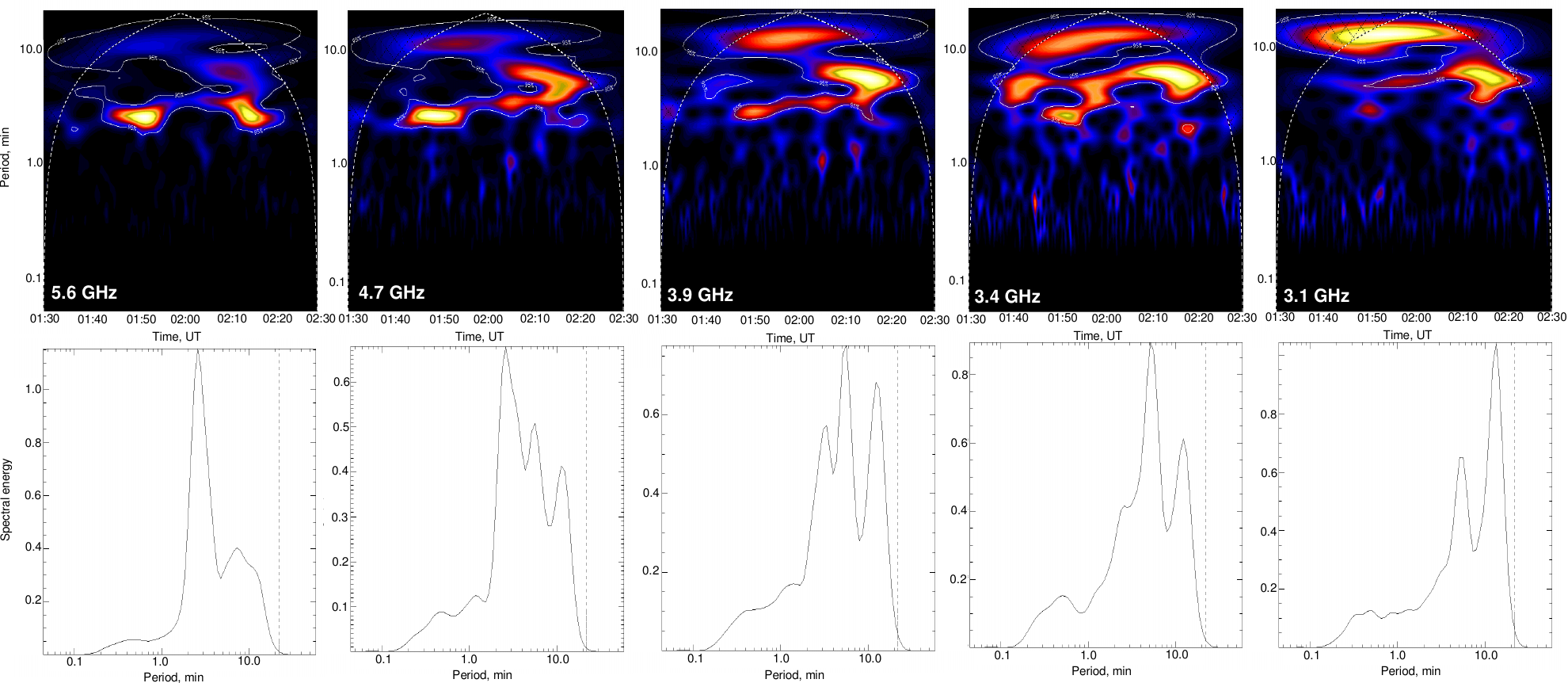}
\caption{Top panels: wavelet power spectra of correlation curves in the intensity channel obtained for 5.6, 4.7, 3.9, 3.4, and 3.1 GHz. Contours show the spectrum regions with 95\% significance level. Bottom panels: Relevant wavelet global spectra.}
\label{6}
\end{figure*}

Figure~\ref{5} shows the results. One can see the signal periodicity on the curve of correlation coefficients shown in brown. In profiles of correlation and radio flux, there is a match between some pulses shown with the blue curve in Fig.~\ref{5}(a). We should note the high level of high  frequency (HF) noise both in the source and signal from the quiet Sun. The average modulation on correlation curve is about 7.4\% with peaks reaching 22\%. For the flow, this value is an order of magnitude less than 0.5\% and peaks of about 1.4\%. In both cases, the maximum modulation is reached at the maximum of oscillations train.

In the region of quiet Sun, mainly HF noise is detected. For numerical comparison of flux variations in the active and quiet regions, we calculated the linear cross-correlation with the correlation curve (see Fig.~\ref{5}b). In the first case, the cross-correlation is positive and reaches a significant level of 0.6, in the second case it is negative and insignificant. The results allow stating that the variations observed on correlation curves are associated with the brightness changes in AR. Further we shall use them to investigate into characteristics of oscillations at different radio frequencies.

\subsection{Spectral characteristics of oscillations}

For spectral visualization of the revealed periodicities in correlation curve signals, we used wavelet analysis \citep{1998BAMS...79...61T}. For each radio frequency in the intensity and polarization channels, wavelet spectra were calculated. Figure~\ref{6} shows the result for the intensity channel in the form of oscillation power spectra (top panels) with their wavelet global spectra (bottom panels). Contours show the levels of oscillation harmonics with 95\% significance level. 

\begin{figure}
\includegraphics[width=8.5 cm]{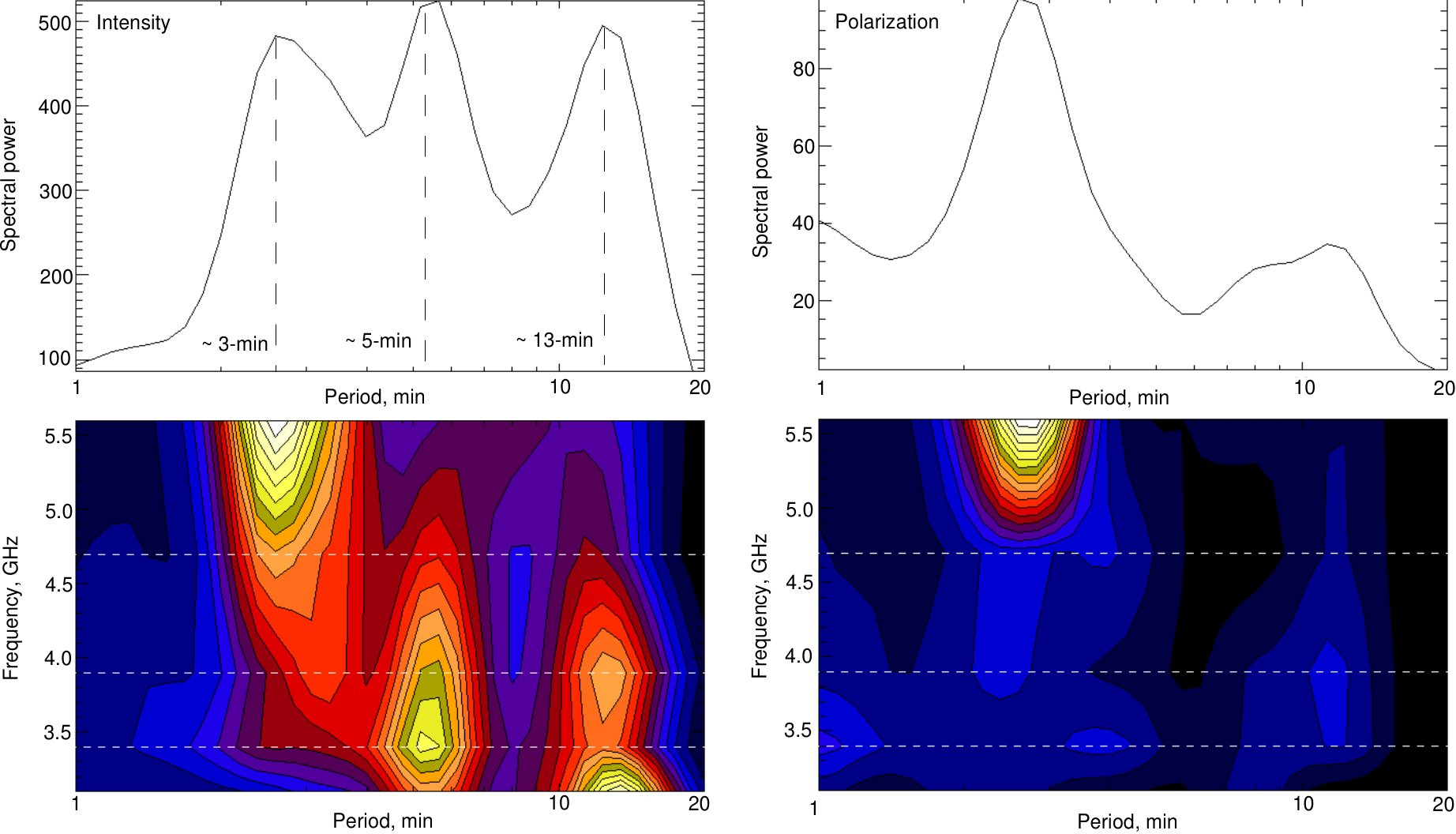}
\caption{Wavelet power spectra of the intensity variation in correlation curves within 3.1-5.6 GHz (bottom panels) and their global wavelet spectra (top panels). Vertical dashed lines depict significant peaks of oscillations.}
\label{7}
\end{figure}

It can be seen that for high radio frequencies of $\sim$5.6 GHz, the main periodicity corresponds to 2–4 min band with the peak of 2.6 min. It forms two wave trains with $\sim$11–14 min duration, which are clearly visible on the correlation curve as oscillations in Fig.~\ref{1}. There are also oscillations in the period ranges of $\sim$5 and $\sim$13 min, but their level is insignificant. At this frequency, HF noise is minimal.

As the receiving frequency decreases, we can see enhancement of the power of long-period oscillations. Gradually, as the power of 3-min oscillations decreases at 4.7 GHz, harmonics begin to enhance with 5-min and 13-min periodicity. On the wavelet power spectra Fig.~\ref{6}, top panels, these changes are seen as smooth brightening of some details with their position changing over the period. The relative power of different harmonics on the global wavelet spectra in Fig.~\ref{6}, bottom panels also changes with the radio frequency.

Comparison of harmonics in the frequency band showed that at all radio frequencies of the intensity channel (see Fig.~\ref{7}, left panels), one can see the revealed three peaks related to 3, 5 and 13 min oscillation periods. Their peaks of power are located at different radio frequencies. Thus, for the 3-min component, the maximum is localized at high frequencies near 5.6 GHz. As the frequency decreases, its power gets lower. There is also a slight drift towards low frequencies over the period. In contrast, for 5-min and 13-min long-period components, we can observe enhancement in the power oscillations reaching their maximum at 3.4 GHz and 3.1 GHz, respectively. Concurrently, the level of HF noise increases. 

In the polarization channel, Fig.~\ref{7}, right panels, there is only one peak in the frequency band related to the 3-min component. The maximum values of oscillation power are seen at 4.7 and 5.6 GHz. The level of long-period oscillations is significantly lower as compared to the intensity channel. The 5-min spectral component has almost completely disappeared. It can be assumed that the radio source spatial structure has different narrowband components that are possibly related to three branches of the obtained periodicity. 

\subsection{Spatial structure of active region and oscillation sources}

As mentioned above, on the correlation curve one can distinguish oscillations with three different periods, whose relative power depends on the radio frequency. All three oscillation branches are caused by variations in radio emission in NOAA 12833 active region. It can be assumed that the observed features of the dynamic spectrum are related to spatial heterogeneity of the oscillation sources at different frequencies.

\begin{figure}
\includegraphics[width=8.5 cm]{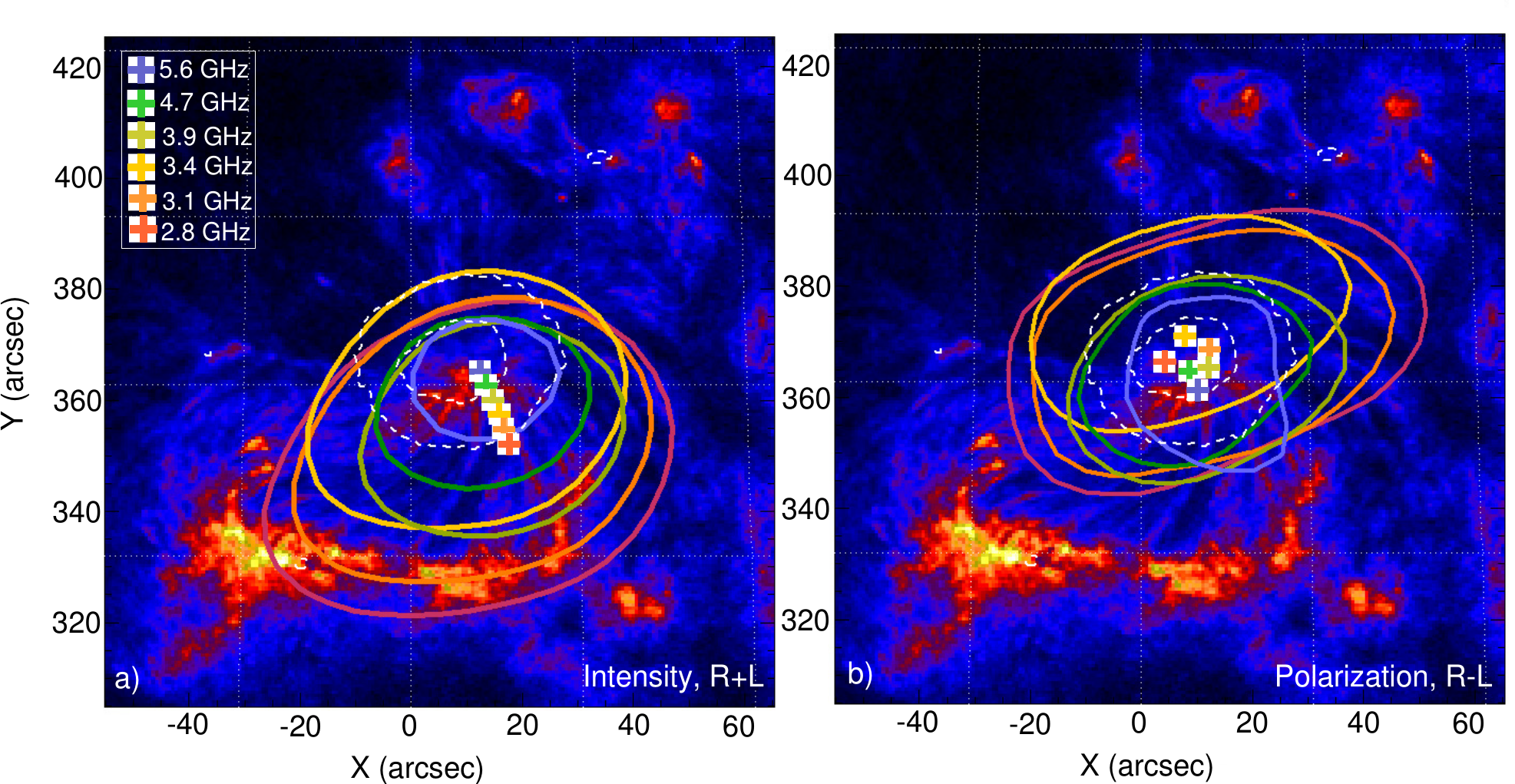}
\caption{(a) Radio images of sunspot active region at 2.8, 3.1, 3.4, 3.9, 4.7 and 5.6 GHz in the intensity channel at 01:30 UT. Colored contours show radio brightness at 0.5 of the maximum for each frequency. A set of crosses show the linear shift of the source maxima at different frequencies. (b) Same for circular polarization channel. SDO/AIA image at 304 \AA ~is the background image.}
\label{8}
\end{figure}

To compare the shape of the sources, their images were plotted in the intensity (R+L), Fig.~\ref{8}(a) and circular polarization channels (R-L), Fig.~\ref{8}(b) at 01:30 UT. Sources at different frequencies are shown as color contours at 0.5 of the maximum radio brightness at relevant frequency. Umbral and penumbral boundaries in the optical range (SDO/HMI, continuum) are shown with two dashed lines.

We can see that in the intensity and circular polarization channels, the sources differ both in shape and location relative to the sunspot center. In the intensity channel (see Fig.~\ref{8}a), the contours are not symmetrical, their maxima are at different distances from the umbra. Spatial shape changes with the frequency. At the frequency of 5.6 GHz, the source is almost symmetrical and coincides with the sunspot. Its maximum brightness is located in the umbra and shown with a cross. As the frequency of radio emission decreases, we can observe a linear trend of the source maxima shifting towards south, where there is a loop structure visible in UV 304 \AA. ~At this wavelength, the height where EUV emission is formed in the transition zone is close to the heights were where radio emission is generated \citep{2012SoPh..275...17L}. For the frequency of 2.8 GHz, the maximum is on the penumbra boundary. At the same time, the shape of sources is changing. We can see their asymmetry increasing as they expand in the southern direction and involve both the loop region and the bright region associated with the pores.

In the circular polarization channel (see Fig.~\ref{8}b), spatial localization of all sources above the sunspot is almost the same. As the frequency gets lower, their angular size increases quasi-symmetrically, forming a spatial funnel centered in the umbral region, where the brightness maxima converge. The size of the sources is slightly larger than the size of the sunspot itself, taking the account of changes in the SRH diagram with frequency. This configuration differs from what we see in the intensity channel.

\begin{figure}
\includegraphics[width=8.5 cm]{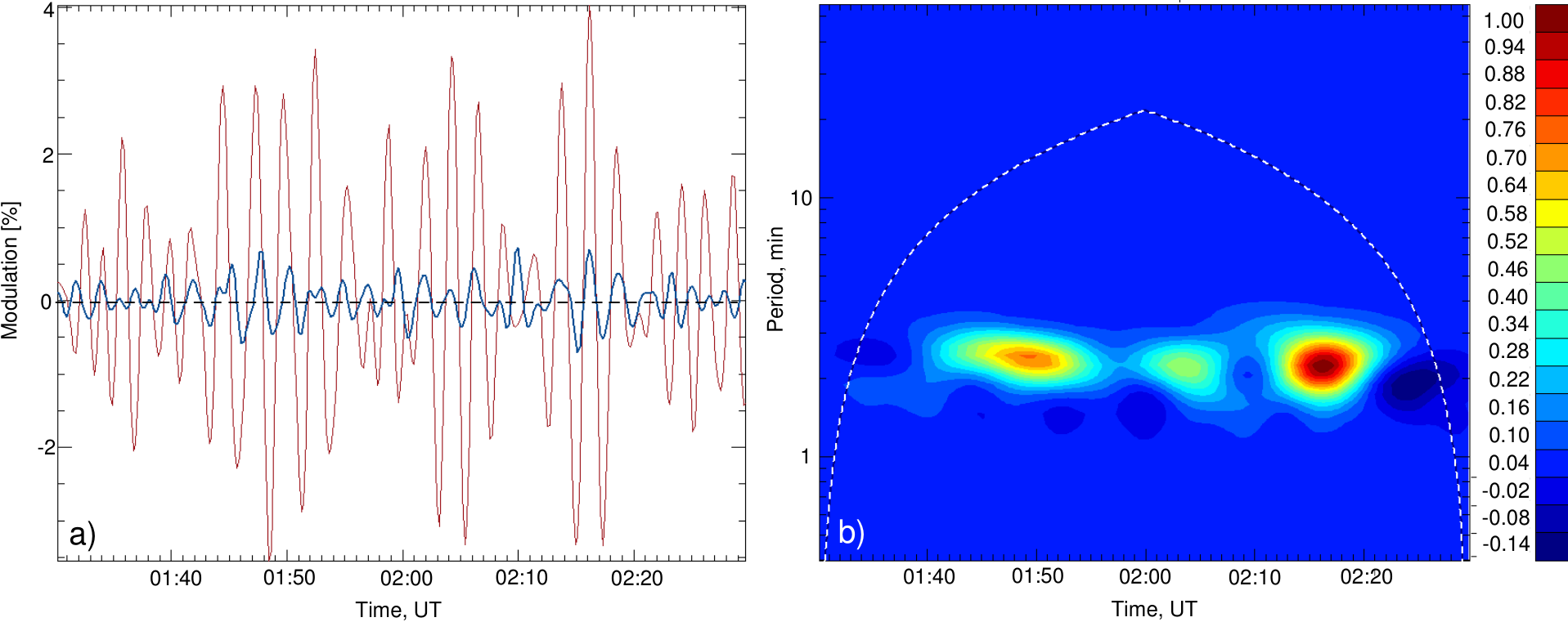}
\caption{(a) Modulation of filtered fluxes in 3-min band of the sunspot UV (SDO/AIA 304 \AA, ~brown curve) and radio (SRH 4.7 GHz, blue curve) emission in NOAA 12833 active region. (b) Wavelet cross-correlation between two curves.}
\label{9}
\end{figure}

UV images show that the radio contours involve many elements of fine structure in these images. Therefore, oscillations might appear on the correlation curves if oscillations of compact UV sources are in-phase at scales comparable to the sizes of radio sources. To verify this statement, we compared the UV emission flux filtered in 2–4 min range (SDO/AIA, 304 \AA) ~integrated over the umbra region and the radio flux time profile in the intensity channel selected in the same range at 4.7 GHz, Fig.~\ref{9}. At this frequency, the maximum radio brightness was near the sunspot central part. This made it possible to compare oscillations in observational data of independent astronomical instruments.

Figure~\ref{9}(a) shows the studied modulation profiles. We can see that the level of UV flux modulation (brown lines) is almost five times higher than that in radio (blue curve). For radio data, the averaged signal variation was 0.3\% with the maximum of 0.7\%. For UV emission, these values were 1.5\% and 4.0\%, respectively. To obtain information about parameters of the correlation of oscillations, we calculated the wavelet cross-spectrum (see Fig.~\ref{9}b) with smoothing of UV images, take into account the different angular resolution of instruments. 

Pronounced are three regions of cross-correlation in 3-min band with 8–13 min duration and coefficient levels varying from 0.5 to 1. From the temporal dynamics of profiles and cross spectrum it follows that 3-min oscillations are modulated by low-frequency wave trains, due to which signals in radio and UV ranges are phased. Positive correlation prevails. Weak negative correlation appears only on the boundaries of the significant interval shown as a dotted curve in Fig.~\ref{9}(b).

\begin{figure}
\includegraphics[width=8.5 cm]{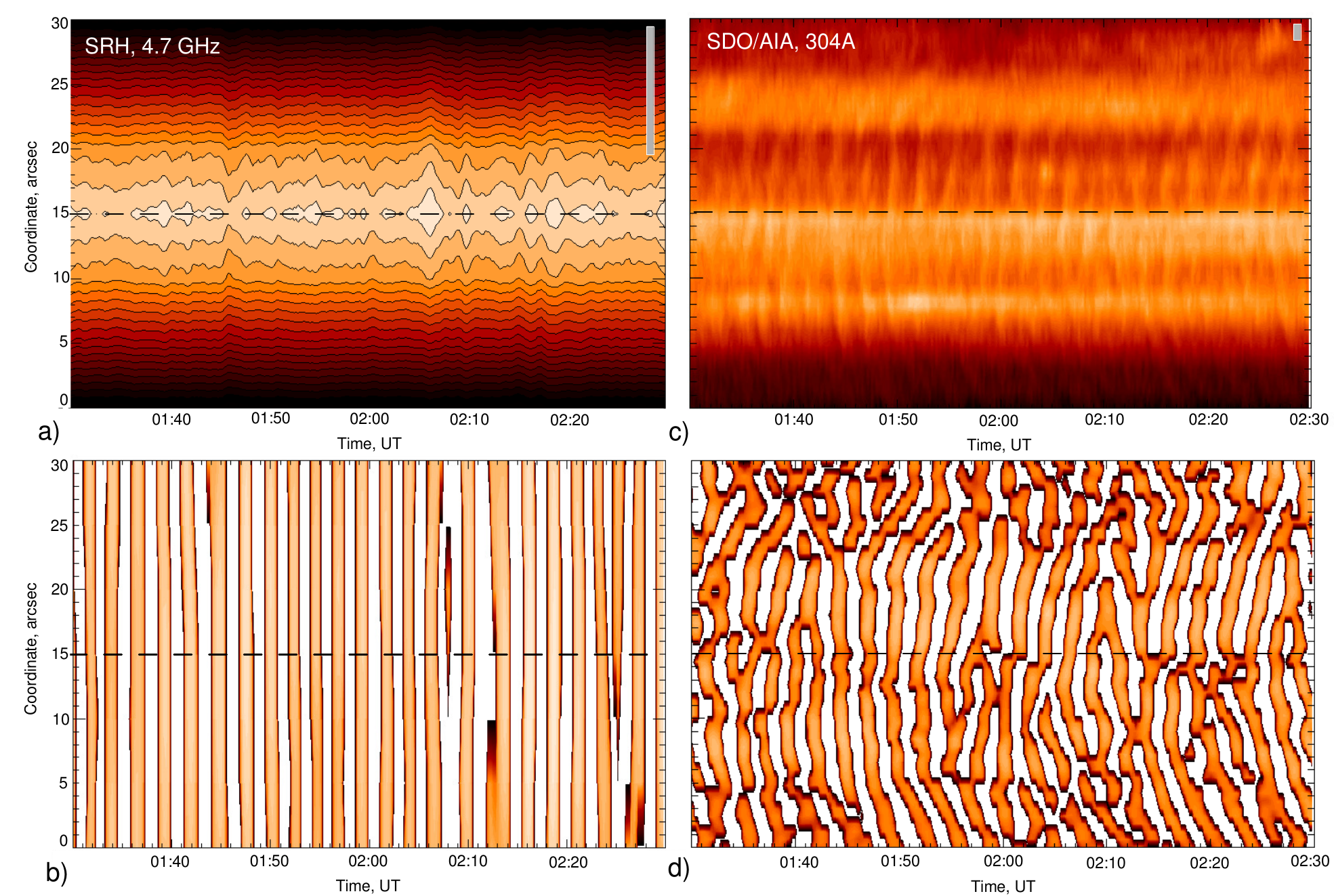}
\caption{Time-distance diagrams in radio (SRH, 4.7 GHz, the intensity channel) and UV (SDO/AIA, 304 \AA). ~Top panels show the original data, bottom panels –filtered data with 3-min periodicity. Horizontal line shows the central part of the source that coincides with the umbra. Gray box shows the size of 1D diagram.}
\label{10}
\end{figure}

To study the structure of wavefronts, we calculated the time-distance diagrams of emission variations, with cross-sections along the x-axis of the source central part at all frequencies in radio and UV. Figure~\ref{10}, top panels show the obtained diagrams at 4.7 GHz frequency and 304 \AA ~wavelength. Horizontal line depicts the central part coinciding with the umbra, where we can see the pronounced $\sim$3 min periodicity. When moving away from the center, the oscillation amplitude decreases.

To identify the wavefront boundaries in diagrams, we used spectral filtering of signals in 3-min band. Bottom panels in Fig.~\ref{10} present the result. We can see differences in the 1D spatial structure of propagating fronts. In the radio band, they are linear             (see Fig.~\ref{10}b). One cannot observe propagation of disturbances in the form of repeated quasi-spherical fronts that are clearly visible in UV emission in Fig.~\ref{10}(d). In UV, the fronts are characterized by a complex shape induced by interaction of different parts of the fronts with different periods and a slowdown in the propagation velocity with distance from the umbra center. In radio, oscillations almost simultaneously involve the entire source and appear like standing waves with a certain periodicity. This dynamics is typical for all wavelengths in 3–6 GHz band and involves whole a source with increasing angular size on frequency. In terms of dynamics, these oscillations are close to what is observed at the photospheric level, where the entire atmosphere above the sunspot experiences similar simultaneous disturbances.

Comparing spatial distribution of the sunspot brightness in different frequency ranges, Fig.~\ref{8}, and variation map of the radio source  Fig.~\ref{3}(b), one can see that signal oscillations in a wide range of periods are distributed throughout the entire active region, not only concentrated in the sunspot as 3-min oscillations. To understand the nature of oscillation processes, distribution of oscillation sources in space must be found, and they must be positioned to similar sources in the UV range.

\begin{figure}
\includegraphics[width=8.5 cm]{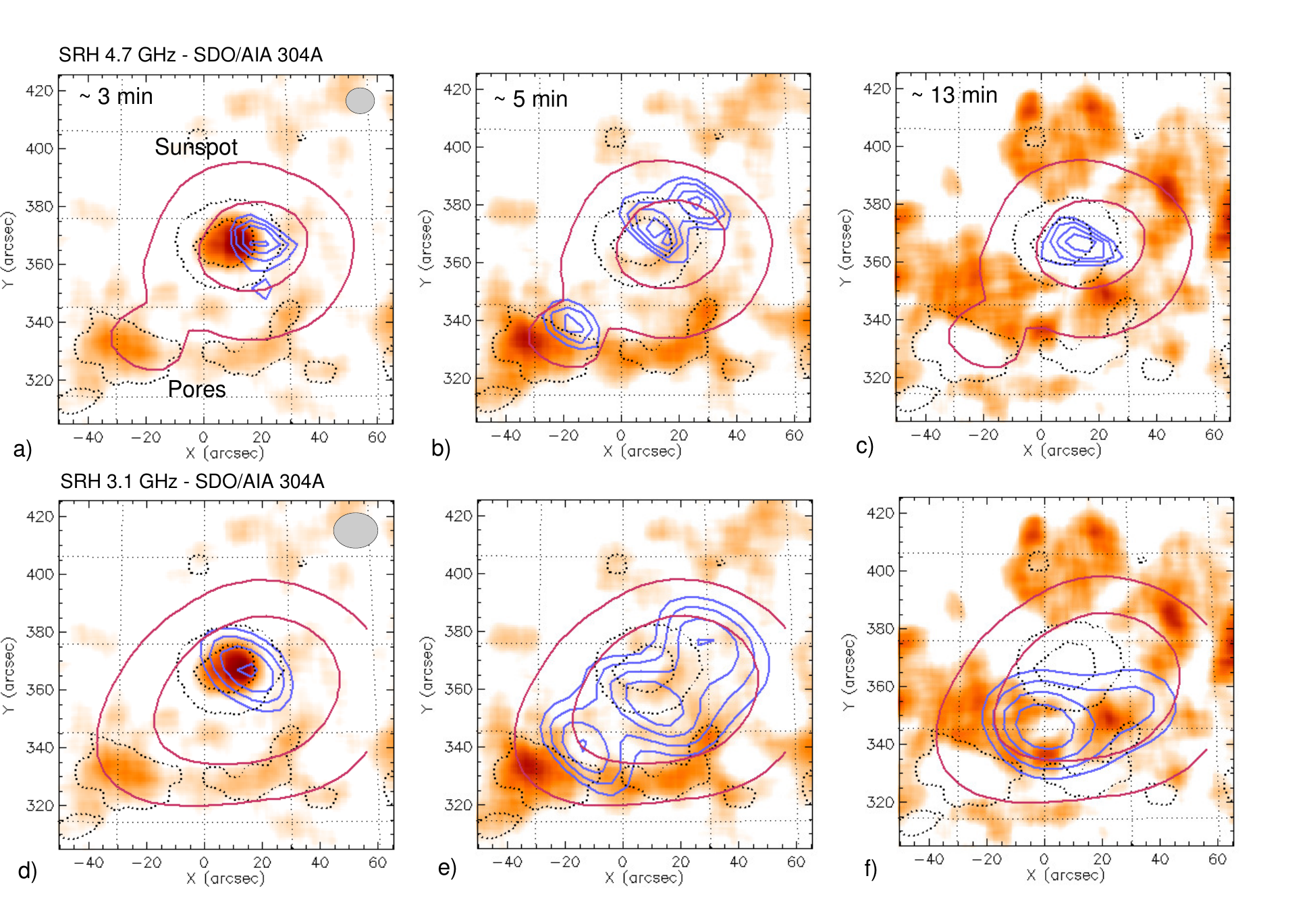}
\caption{Narrowband distribution maps of oscillation power with $\sim$3, $\sim$5 and $\sim$13 min periodicity in UV (SDO/AIA, 304 \AA, ~background) with superimposed blue contours of narrowband radio sources in the intensity channel at 4.7 and 3.1 GHz. Red contours at the levels 0.1 and 0.5 of the maximum emission show snapshots of radio sources at 01:30 UT. Dotted contours indicate the sunspot boundaries (umbra, penumbra) and pore region. The diagram is shown as a gray filled ellipse.}
\label{11}
\end{figure}

We compared narrowband sources of different periods at 3.1 GHz and 4.7 GHz, as well as sources at the wavelength of 304 \AA. ~Time series of variations in amplitudes of different pixels of image cubes were used to obtain maps of oscillations distribution throughout the active region. These data underwent spectral decomposition into elementary harmonics using the pixel-by-pixel wavelet filtering method (PWF analysis) \citep{2008SoPh..248..395S}. The method allows obtaining narrowband signals in the spectrum and computes their dynamics of power distribution in time and space. We calculated oscillation maps for the previously found peaks of oscillation periods of $\sim$3, $\sim$5 and $\sim$13 min. For the study, we selected the maps obtained during the maximum oscillations for each periodicity: 01:48 UT (3 min), 02:18 UT (5 min) and 02:00 UT (13 min). At this time, the sources of oscillations are most clearly visible. To compare maps in radio and UV with different spatial resolutions, we smoothed UV images. Figure~\ref{11} shows the result.

We can see in Fig.~\ref{11}(a), background that 3-min periodicity in the UV range is mainly localized within the umbra boundaries depicted with dotted contours. The oscillation power is maximal here. Also, there are different minor sources in the form of small spots around the main sunspot in the superpenumbral region. According to oscillation power, a semiring is distinguished that involves the sunspot and pores of the opposite polarity in the south. The amplitude of 3-min oscillations in the semiring is considerably lower than in the sunspot. When the oscillation period increases to 5 min, the umbral bright source disappears and the semiring structure on the umbra/penumbra boundary becomes the brightest structure (see Fig.~\ref{11}b). Sources of 13-min oscillations are bound to the arcade of low loops coupling the sunspot with the pore region. In this superpenumbral region, narrowband sources are shaped as extended loops in Fig.~\ref{11}(c). Basically, they almost completely encircle the sunspot, except the northeast direction, where high magnetic loops are seen in 171 \AA ~coronal line     (see Fig.~\ref{2}f).

Positions of the narrowband oscillation sources in the radio range (blue contours) are in good agreement with similar sources in the UV range. Red contours in Fig.~\ref{11} outline the boundaries of snapshots of radio sources at 01:30 UT at the levels 0.1 and 0.5 of the maximum brightness.  Two radio sources are distinguished; one is related to the sunspot and the other – to the region of the maximum pore brightness. The 3-min narrowband sources shown with blue circles at 4.7 GHz and 3.1 GHz in Figs~\ref{11}(a) and (d) demonstrate good agreement only with the umbra. They are not found beyond the sunspot. Their brightness levels are 0.5–0.9 of the maximum brightness. The obtained match of localization of 3-min sources at both frequencies with the sunspot and maximum UV oscillations evidences a single mechanism of disturbances in different ranges.

For 5-min period, positions of three narrowband radio sources are associated with the outlined regions on the umbra/penumbra boundary in Figs~\ref{11}(b) and (e) and in the pore region where oscillations in UV range demonstrate the power enhancements with similar periodicity. It should be noted that in this range of  observed periods are not extended structures as in UV, but compact sources. Their positions mainly coincide with footpoints of magnetic field lines and their brightening in the local.

The part of the spectrum with the longest period as 13-min oscillation sources in the radio range is represented in different ways, at different frequencies. For 4.7 GHz, the narrowband source is compact and matches the umbra. For low frequency of 3.1 GHz, in the superpenumbral region, we can see an arc-shaped extended source spatially distant from the sunspot. Here, above the arcade of low magnetic loops, is the maximum of oscillations in UV as a bright semiring. The radio source involves only a part of this region.

\section{Discussion}

We studied a bright polarized source observed in the 3–6 GHz band using data from SRH,. It was above a large sunspot of NOAA 12833 active region. Its temperature ranged from 1.3 to 1.9 x $10^6$ K at 2.8–4.7 GHz. Polarization was changes from 5.2\% to 12.1\%. This means that the emission was mainly generated by thermal electrons on the gyrofrequency third harmonic, in the layer of the sunspot atmosphere with $\sim$2000–3000 G magnetic field. Also, the active region has a peripheral part in the form of pores with $\sim$1000 G field, where the mechanism of thermal free-free emission prevails.

The solar disk map in Fig.~\ref{3} shows that the source of the maximum variation in radio emission coincides with the active region. There are three factors influencing the occurrence of a certain periodicity of signals from the sunspot-associated sources. The first is related to their spatial structure in the radio range at different heights \citep{2002A&A...386..658N}. The second factor is related to the chromospheric resonator in the umbra and physical characteristics of the medium \citep{2014AstL...40..576Z}. The third factor is the mechanism of frequency cutoff of slow magnetoacoustic waves propagating as traveling waves in penumbra \citep{2012ApJ...746..119R, 2020ApJ...888...84S}. All three factors work at a time, so observations with high spatial resolution are essential to understand the quantity of their impact. Also, high time resolution is important to detect different harmonics in a signal. We used data from the Siberian Radioheliograph in 3–24 GHz band and SDO/AIA/HMI spacecraft, which possess these characteristics.

We suppose that the detected oscillations in the radio range with $\sim$3, 5 and 13-min periodicities are associated with oscillations in sunspots previously found at 17 GHz (NorH) \citep{1999SoPh..185..177G, 2001ApJ...550.1113S, 2008SoPh..248..395S}, 5 GHz and 8.5 GHz (VLA)    \citep{2002A&A...386..658N}, and in the optical and UV ranges \citep{2000ApJ...534..989B, 2007PASJ...59S.631N, 2009Ge&Ae..49..935K}. This conclusion is based both on corresponding periods found and on temporal dynamics of waves in the form of low-frequency trains that exist throughout the observation time.

We obtained a direct correspondence between the sunspot passage across the solar disk and the increase in the apparent power of 3-min oscillations in radio near the central meridian (see Fig.~\ref{1}b). The region with the observed enhancement in radio signal variation is limited to 40 degrees of heliolongitude. Outside this zone, oscillations are reduced due to the projection effect and active region evolution. Intensity variations and non-symmetry of the obtained curve can be explained by changes in the magnetic field structure and its projection toward the observer. The umbra shape and area in NOAA 12833 group changed as it moved from the east to the west limb. At this time, new pores formed near the sunspot, reaching their maximum near the center. Since 3-min oscillations prevail in the sunspot umbra, we can see that on the limbs, its area is the lowest and, accordingly, minimum emission variations can be observed. On the central meridian, the sunspot has the maximum umbral area with the vertical field pointing strictly at the observer. Accordingly, 3-min variations will be maximum here.

We studied one-hour data interval in radio and UV with high time resolution. This made it possible to identify periods in the range from 0.1 to 20 minutes using data spectral analysis. Global wavelet spectra of correlation coefficients (see Fig.~\ref{6}) obtained at different frequencies exhibit concentration of oscillation power of $\sim$3, 5 and 13 min periods. We do not see multiple harmonics of the 3-min period in the spectra. The existence of the found 3-min peak in the radio range is in good agreement with the similar periodicity observed earlier at 17 GHz, and with its dynamics in the form of low-frequency wave trains \citep{2001ApJ...550.1113S, 2012A&A...539A..23S}. The detected 5-min component was previously noted on the power spectra in \cite{1999SoPh..185..177G}, but its level was small. A possible explanation for this is insufficient time interval, and the emission frequency, at which sources with narrowband periodicity have small size and, accordingly low level of oscillations. The low-frequency 13-min peak is presumably associated with low-frequency waves that exist both in the sunspot and superpenumbral region.

Time delays between 3-min oscillations at different frequencies indicate that their sources are at different heights in the umbral atmosphere. The revealed positive difference of phases between high and low frequencies is related to the upward propagation of waves along the vertical field lines anchored in the umbra. Penetrating high into the atmosphere above the sunspot, the waves successively modulate the gyroresonance levels, at which radio emission occurs, which leads to the observed periodicity and occurrence of signal delay. The observed phase changes during the propagation of trains of 3-min oscillations that occur for individual pairs of radio frequencies can be interpreted by the asymmetry of individual pulses with the occurrence of frequency drifts \citep{2012A&A...539A..23S}. 

The one-dimensional spatial structure of 3-min wavefronts in the radio range showed the presence of standing waves in 3–6 GHz band in Fig.~\ref{10}(a). No changes are seen in the wavefronts from the sunspot center as in UV, Fig.~\ref{10}(c). Oscillations involve the entire source simultaneously. The fronts are parallel to each other and have sharp edges. One can observe similar oscillation dynamics at the photospheric level in the umbra with the vertical magnetic field. We assume that simultaneous changes in the brightness of the sources are caused by periodic disturbances of the 3rd level of gyrofrequency by propagating waves. Since this layer is narrow and its $\sim$100 km (17 GHz) \citep{1999SoPh..185..177G}  or $\sim$25 km (5 and 8 GHz) \citep{2002A&A...386..658N} shifts in height are insignificant compared to the atmosphere altitude above the sunspot, the observer will see these oscillations as standing waves.

Maps of narrowband oscillations in Fig.~\ref{11} showed that 3-min oscillations prevail in the intensity channel at 3.1 and 4.7 GHz. Their sources are above the umbra in the optical range and in the maximum of UV oscillations. In the polarization channel, all maxima of radio sources are within the umbra with 3-min oscillations. The maximum oscillations is observed near 5.6 GHz. The increase in the size of 3-min source at low frequencies can be explained by the increased tilt of magnetic field lines within the umbra, along which waves are propagating. Although these changes are weak, given the quasi-verticality of the field, they can increase the size of resonator and, accordingly, gradual decrease in the oscillation frequency. Recent studies \citep{2020NatAs...4..220J, 2021NatAs...5....2F}  showed formation of a similar spatial funnel in the umbra, where the researchers observed resonant amplification of signals.

Propagating waves can affect the environment and induce oscillations in density and temperature. This, in turn, causes 3-min oscillations in the wave cutoff frequency and corresponding changes in the radio brightness of the source. Earlier data showed \citep{1999ApJ...517L.159B} that relative oscillations in density of $\sim$5\% and temperature of $\sim$3\% can cause oscillations in radio brightness of $\sim$14\% at 17 GHz, assuming optically fine gyroresonance emission at the level of the gyrofrequency third harmonic. It is close to the observed modulation of the correlation coefficients $\sim$7\%, taking into account the lower frequency of 4.7 GHz, where the amplitude of high-frequency oscillations goes down.

In \cite{2002A&A...386..658N} it is shown that the observed microwave oscillations in the sunspot can be explained by strength oscillations of the magnetic field. This leads to changes in the levels of the second and/or third harmonics of the gyrofrequency in height with $\sim$25 km amplitude. We know that in the sunspot region, the plasma beta is significantly lower than unity, which prevents propagating waves from changing the curvature of the field lines in transverse direction. We can assume that the visible changes in fronts' symmetry during the development of 3-min wave trains evidence the occurrence of shock waves at this time. Also, such changes in the fronts are related to the occurrence of frequency drifts \citep{2012A&A...539A..23S}. All this can increase local pressure and plasma beta in the vicinity of the propagation region and, accordingly, cause magnetic field disturbances with a certain periodicity. Earlier, similar field disturbances were seen during propagation of traveling waves in penumbra \citep{2013A&A...556A.115D} and appearance of umbral flashes \citep{2018ApJ...860...28H}. 

When considering 5-min oscillations, we can see in Fig.~\ref{11} changes both in the number and location of the sources. In UV, brightness in the umbra fades out abruptly to form a ring-like detail on the umbra/penumbra boundary. In \cite{2003MNRAS.346..381C, 2006MNRAS.372..551S}~similar localization of 5-min narrowband sources was interpreted as indication of a strong interaction between acoustic waves and the magnetic field in the region where the local conditions favor the absorption of the global solar p-mode. 

Brightness in the pore region has also increased. In radio, these changes are due to the appearance of three discrete sources near these new formations. Poor angular resolution of SRH compared to SDO/AIA makes it impossible to identify all fine-structure details of these objects. Their position is presumably due to the peculiarity of magnetic field, where we can see concentration of magnetic waveguides with the maximum brightness of emission in Fig.~\ref{2}. It is this region where the maximum oscillations are observed. For all the revealed sources, they are sinphased. 

The obtained results are in good agreement with data obtained earlier at 17 GHz \citep{2008SoPh..248..395S}. It was shown that narrowband sources of oscillations have different spatial distribution over the sunspot. Located in umbra are the sources of 3-min and 15-min oscillations, while the sources of 5-min oscillations belong to small regions on the umbra/penumbra boundary. Similar spatial distribution of oscillations in sunspots was observed in \citep{2000A&A...355..347Z} and \citep{2002A&A...386..658N}. 

In contrast to distribution of 3 and 5-min oscillations, sources of 13-min waves show different spatial locations at 3.1 and 4.7 GHz. While the source is in the sunspot at 4.7 GHz, at 3.1 GHz it is above the loop arcade, at the distance of $\sim$20\arcsec ~from the sunspot. There is $\sim$130 s delay between signal arrivals at different frequencies. A similar delay was also obtained between 3.1 GHz and 304 \AA ~in UV. All ranges are characterized by sinphased waves, which indicates a common source of disturbances.

The measured time delay enabled us to calculate the waves propagation velocity at different frequencies. It reaches $\sim$100 km/s, which is higher than the velocity of 10–70 km/s measured in \cite{2021RSPTA.37900183M} for 10-min waves, but it coincides with the 40–150 km/s range for 3-min waves \citep{2021ApJ...914L..16C}. It can be assumed that low-frequency trains of 3-min oscillations are modulated by $\sim$13-min waves that can propagate beyond the sunspot into the superpenumbral region. In \cite{2022ApJ...933..108C}, the authors showed such low-frequency Alfven waves with $\sim$10 min period that are excited by convective motions of granules at the photospheric and subphotospheric levels. Such periodicity can be caused in sunspots with strong magnetic field in the form of magnetic convection. While propagating upwards into the corona, initial disturbances in the form of umbral kink waves at the photospheric level convert into transverse Alfven waves beyond the sunspot with 3-min component modulation inside the umbra.

It was shown in \cite{1977A&A....55..239B} that long-period waves can penetrate the photosphere/chromosphere layers and propagate upward into the transition zone and corona. The magnetic structure of the active region with sunspots and surrounding pores forms natural waveguides. Modification of the inclination of these magnetic waveguides and corresponding changes in the wave cutoff frequency can explain the occurrence of low-frequency oscillations in chromospheric spicules \citep{2004Natur.430..536D}, coronal loops \citep{2002A&A...387L..13D, 2005ApJ...624L..61D} and peripheral parts of flare sites \citep{2009ApJ...702L.168D}. Observations of long-period oscillations in the corona have been also explained by inclined magnetic channels along which the waves propagate \citep{2009ApJ...697.1674M, 2011SoPh..272..101Y}.

We assume that the revealed 13-min oscillations at 3.1 GHz above the loops arcade are associated with the gyrofrequency-level modulation by low-frequency Alfven waves propagating from the sunspot into the superpenumbral region. Their phasing with the same periodicity found in the sunspot at 4.7 GHz indicates the same waves and the source of disturbances. The different localizations of the oscillation sources at these frequencies can be explained by the modulation of gyroresonant layers by waves with different cutoff frequencies. At high frequencies, this spectral component is in the umbra with a vertical field and is connected with the modulation of 3-min oscillations in the form of wave trains. At low frequencies, it is located beyond the sunspot above the inclined low loops.

\section{Conclusions}

The SRH spatial and temporal resolution allows one to observe quasi-periodic oscillations of microwave emission from the sunspot active region NOAA 12833 with 2–13 min periods. For the first time, the spatial resolved observation of these oscillations were obtained simultaneously in the intensity and polarization channels at several frequencies in 3–6 GHz band and compared with UV observations. The main results can be presented as follows:

We revealed the changes in the observed power of 3-min oscillations during the passage of NOAA 12833 active group across the solar disk. The highest power is observed in the central region of the Sun,https://www.overleaf.com/project/63319ecb79bdc5abb377d84b bounded by heliographic longitudes of $\pm$40 degrees, with the maximum near the central meridian.

Spectral analysis of radio emission shows the peaks corresponding to the periods of $\sim$3, 5 and 13 min. It is found that 3-min component reaches its maximum power at high frequencies near 5.6 GHz. As the oscillation period increases, the peak power in the spectrum shifts towards low radio frequencies. There is only a 3-min peak near 5.6 GHz in the polarization channel.

The dynamics of 3-min oscillations in radio shows modulation in the form of low-frequency wave trains. We found a positive time delay between periodic pulses, which increases to low radio frequencies. This attests upward propagation of oscillations.

Comparison of the sunspot emission dynamics in the radio (4.7 GHz) and UV (304 \AA) ~bands shows the positive cross-correlation. The maximum coefficients are near the 3-min period and correlate with the time development of wave trains. This indicates the proximity of the heights of the studied periodicity sources. Lower spectral periodicities do not correlate well with each other.

The structure of 3 min wavefronts in the radio band shows the standing waves. They involve practically the all sunspot umbra simultaneously and can be explained by periodic disturbances of gyroresonance layers in height above the sunspot. 

For the first time, we detected spatial localization of narrowband sources of microwave oscillations from the active region in 3–6 GHz band. It is shown that the major oscillations are related to 3–min polarized sources located above umbra and coincided with the place of maximum oscillations in the UV range. Location of these sources in the intensity channel coincides with that in the polarization channel. At lower radio frequencies, the level of 3-min component drops abruptly. Sources of 5-min oscillations concentrate on the umbra/penumbra boundary and near the pore region. Narrowband sources in the radio and UV bands coincide spatially. It is found that locations of 13-min low-frequency component are different at different radio frequencies. At high frequency of 4.7 GHz, it coincides with the sunspot, at low frequency of 3.1 GHz it is connected with the low loop arcade binding the sunspot and pores.

We assume that the revealed narrowband sources of oscillations are related to different magnetic waveguides, along which propagating waves with different periodicities and oscillation modes modulate the gyrofrequency layers. Spatial heterogeneity of oscillation sources in active region is formed due to a resonator of broadband signals in umbra and cutoff frequency in penumbra.

The obtained results show the presence of significant oscillations in the radio range and their unambiguous relationship with non-stationary processes observed in UV. These common processes reflect propagation of slow magnetoacoustic waves from the subphotospheric level into the corona along tilted magnetic fields from the sunspot. The obtained wave power dependence on the oscillation period is determined by spatial location of magnetic waveguides, along which disturbances propagate with modulating corresponding gyrofrequency levels in the radio range.

\section*{Acknowledgments}

We are grateful to the teams of the Siberian Radioheliograph and Solar Dynamic Observatory, who have provided access to their data. We thank Dr. S.A. Anfinogentov for his assistance in processing the experimental data and Dr. Alexey Kuznetsov for the fruitful discussion. Development of the methods used in Sect. 2 was supported by the budgetary funding of Basic Research program No. II.16. This work was supported by the Ministry of Science and Higher Education of the Russian Federation, and the Russian Science Foundation (Grant No. 21-12-00195).

\section*{Data Availability}

Experimental data were obtained using the Unique Research Facility Siberian Solar Radio Telescope \footnote{\url{http://ckp-rf.ru/usu/73606}}, and the equipment of the Shared Equipment Center “Angara” \footnote{\url{http://ckp-rf.ru/ckp/3056}}.  The SDO data can be accessed from the Joint Science Operations Centre \footnote{\url{http://jsoc.stanford.edu}}. The Solar Soft library located on \footnote{\url{https://sohoftp.nascom.nasa.gov/solarsoft/}}. 

\bibliographystyle{mnras}
\bibliography{BibTex_Sych} 


\bsp	
\label{lastpage}
\end{document}